\documentclass[12pt]{iopart}
\usepackage{amssymb}
\usepackage{bm}
\usepackage{graphicx}
\usepackage{epstopdf}
\expandafter\let\csname equation*\endcsname=\relax
\expandafter\let\csname endequation*\endcsname=\relax
\usepackage{amsmath}
\newcommand{\be}{\begin{equation}}
\newcommand{\en}{\end{equation}}

\newcommand{\pre}{Phys. Rev. E }

\newcommand{\avg}[1]{\left< #1 \right>}

\usepackage{tikz}
\usetikzlibrary{calc}

  \makeatletter
\newcommand{\vast}{\bBigg@{1}}
\newcommand{\Vast}{\bBigg@{3}}
\newcommand{\Vvast}{\bBigg@{4}}
\makeatother
\begin{document}
\title{Statistical properties of eigenvectors and eigenvalues of structured random matrices}
\author{K. Truong, A. Ossipov}
\address{School of Mathematical Sciences, University of Nottingham, Nottingham NG7 2RD, United Kingdom}

\begin{abstract}

We study the eigenvalues and the eigenvectors of $N\times N$ structured random matrices of the form $H = W\tilde{H}W+D$ with diagonal matrices $D$ and $W$ and $\tilde{H}$ from the Gaussian Unitary Ensemble.  Using the supersymmetry technique we derive general asymptotic expressions for the density of states and the moments of the eigenvectors. We find that the eigenvectors remain ergodic under very general assumptions, but a degree of their ergodicity depends strongly on a particular choice of $W$ and $D$.   For a special case of $D=0$ and random $W$, we show that  the eigenvectors can become critical and are characterized by non-trivial fractal dimensions.

\end{abstract}

\maketitle
%*****************************************************************************************************
\section{Introduction}

Statistical properties of eigenvalues and eigenvectors of random matrices is the central topic of Random Matrix Theory (RMT) \cite{RMT_book} .  The key idea of RMT is that many features of complex systems are universal and therefore they can be modelled by ensembles of random matrices, which share the same global symmetries, but do not contain any system specific information. A prominent example of such classical ensemble is the Gaussian Unitary Ensemble (GUE), in which the only constraint is the Hermiticity of a matrix.    

Despite the great success of classical RMT  during the last fifty years, there is a growing interest to new ensembles of random matrices, in which some structural information about an original system is partly present. In this paper, we study one of  such random matrix models, which is defined as
\begin{equation}\label{model-def}
H = W\tilde{H}W+D,
\end{equation}
where $\tilde{H}$ is an $N\times N$ matrix from GUE and $W$, $D$ are diagonal matrices with elements $w_i$ and $d_i$, $i = 1,...,N$, respectively; the matrices $W$ and $D$ can be either deterministic or random.  Since the presence of the matrices $W$ and $D$ breaks the unitary invariance of the probability distribution of $\tilde{H}$, it is reasonable to expect that  the statistical properties of  the eigenvectors of this model might be very different from the corresponding properties of GUE. How exactly they will be different, is the main question addressed in this work.

The random matrices of the form $H=LJR +M$, where $L$, $R$ and $M$ are not necessary diagonal and the random matrix $J$ might be from another random matrix ensemble, appear naturally in various applications including signal processing \cite{NE08}, vibration analysis \cite{S03}, wireless communication \cite{CD14}  and neural networks \cite{AFM15}.  For example, they arise in the linearized dynamics of non-linear neural networks: $J$ is a random connectivity matrix and $L$, $R$ and $M$ can be expressed through the firing rates and the time constants of the neurons \cite{AFM15}.  In the present work we restrict ourselves to the technically simplest case, where $L=R$ and $M$ are diagonal matrices and $J$ is from GUE. 

The spectral properties of such random matrices have been studied recently and a number of very general results have been derived (see \cite{AFM15,GG16} and references therein), however much less is known about their eigenvectors \cite{BY17}. In this work, we generalize our recent results, which have been obtained for two particular cases: i) $D=0$ and $W$ is deterministic \cite{R1} ii) $W=I$ and $D$ is either deterministic  or random \cite{sumrm}. 

One of the main results of this paper is a general non-perturbative, asymptotic expression for the moments of the eigenvectors of $H$, which allows us to calculate the moments for any given values of $w_i$ and $d_i$. From this expression,  it follows, in particular, that the eigenvectors $H$ remain qualitatively the same as the eigenvectors of $\tilde{H}$ for very generic choice of parameters $w_i$ and $d_i$. That means, that extended nature of the GUE eigenvectors is very robust under a wide class of the deformations described by Eq.\eqref{model-def}. At the same time, it also shows that on a quantitative level the eigenvectors of $H$ can be very different from their GUE counterparts, namely they can  occupy an arbitrarily small fraction of the available space.

Another important conclusion following from the general result for the moments  is that the extended nature of the eigenvectors can be altered, provided that $d_i$ and $w_i$ become $N$-dependent. One of the special cases we study in the present paper is the model with $D=0$ and 
uncorrelated Gaussian distributed $w_i$ with the variance, which is $N$-dependent. Such a model can be considered as a multiplicative counterpart of the  Rosenzweig-Porter model \cite{RP60}, whose eigenvectors statistics was calculated in \cite{sumrm}. We find that eigenvectors of this model can be fractal and compute their fractal dimensions.   

The paper is organized as follows. In Section \ref{sec-moments} we derive our general results for the moments of the eigenvectors and the density of states. In Section \ref{sec_general_formula} we investigate a special case of the model with $D=0$ and random $W$. Finally, some conclusions and open problems are discussed briefly in Section \ref{sec_conclusions}.

%*******************************************************************************************************
%*******************************************************************************************************
%*******************************************************************************************************
\section {Moments of the eigenvectors and the density of states}\label{sec-moments}
In this section we derive  expressions for the moments of the eigenvectors of $H$ and the density of states. Generally, the local moments at  energy $E$ are given by the definition

\begin{equation}
I_q(n)=\frac{1}{\rho(E)}\sum_{\alpha}\avg{|\psi_n^{\alpha}|^{2q}\delta(E-E_{\alpha})},
\end{equation}
where $\psi^{\alpha}$ is a normalized eigenvector corresponding to the eigenvalue $E_{\alpha}$  and $\rho (E)$ is the density of states 
\begin{equation}
\rho(E)=\frac{1}{N}\sum_{\alpha}\avg{\delta (E-E_{\alpha})}.
\end{equation}
The integer moments can be related to the diagonal matrix elements of the Green's functions
\begin{equation}\label{mom-green}
I_q(n) = \frac{\mathrm{i}^{2-q}}{2\pi \rho(E)N} \lim_{\epsilon \to 0}(2\epsilon)^{q-1}\avg{(G^R_{nn})(G^A_{nn})^{q-1}}, \quad q=2,3,\dots,
\end{equation}
where $G^R$ denotes the retarded Green's functions and similarly $G^A$  the advanced Green's function, which are defined by
\begin{equation}
G^{R/A}(E) = (E\pm \mathrm{i}\epsilon -H)^{-1},
\end{equation}
where $\epsilon>0$ provides an infinitesimal imaginary shift of $E$ into the complex plain and $\avg{\dots}$ denotes an average over the random matrix ensemble. For the matrix elements of the Green's functions such an average can be computed  by employing the supersymmetry technique. In this approach the averaged Green's functions are represented as superintegrals over a supermatrix $Q$, which is in our case is just a $4\times 4$ matrix. The first steps of the method are very generic and do not depend significantly on the structure a matrix $H$, therefore we do not present them here, further details of the derivation can be found in \cite{R1}.  The superintegral representing the product of the Green's functions from Eq.\eqref{mom-green}  is given by
\begin{equation}
\begin{aligned}
\left<\left(G^R_{nn}\right)\left(G^A_{nn}\right)^{q-1}\right> =
\int dQ (g_{aa}^{BB})^{q-2}\left\{g_{aa}^{BB}g_{rr}^{BB} + (q-1)g_{ar}^{BB}g_{ra}^{BB}\right\}\\ \times  \exp\left\{-\frac{N}{2}\mathrm{Str}Q^2-\sum_i^N\mathrm{Str}\ln(E-\mathrm{i}\epsilon \hat{\Lambda}-d_i-w_i^2Q)\right\}.
\end{aligned}
\end{equation}
$\Lambda = \mathrm{diag}(1,1,-1,-1)$ and $g^{BB} = (E-d_n-w_n^2Q_n-\mathrm{i}\epsilon\Lambda)_{BB}^{-1}$, where the explicit expression for $g^{BB}$ is given in Appendix A. We notice that the standard action of the superintegral appearing in the GUE case is altered by the parameters $d_i$ and $w_i$ as expected.

In the limit $N \to \infty$, the integral is dominated by the saddle-points that satisfy the saddle-point equation
\begin{equation}
Q = \frac{1}{N}\sum_{i=1}^N \frac{w_i^2}{E-d_i-w_i^2Q},
\end{equation}
where the solutions can be parametrized as \cite{F99}
\begin{equation}
Q_{s.p.} = t+\mathrm{i}sT^{-1}\Lambda T,
\end{equation}
the variables $s \neq 0$ and $t$ are two real parameters satisfying the simultaneous equations
\begin{equation}\label{sim-eqns}
t = \frac{1}{N}\sum_{i=1}^N \frac{w_i^4(E-d_i-w_i^2t)}{(E-d_i-w_i^2t)^2+w_i^4s^2}, \quad 1 = \frac{1}{N}\sum_{i=1}^N \frac{w_i^4}{(E-d_i-w_i^2t)^2+w_i^4s^2}.
\end{equation}
In this way any physical quantity, which can be expressed through the Green's functions, can be calculated in terms of $s$ and $t$ by computing the corresponding superintegral over $T$. Then for any given set of parameters $\{d_i\}$ and $\{w_i\}$ the above system of the equations can be solved numerically yielding an explicit result for any quantity of interest. In particular, one can compute the density of states, which takes  the form
\begin{equation}\label{dos-main}
\rho (E) = \frac{s}{\pi N} \sum_{i=1}^N\frac{w_i^2}{(E-d_i-w_i^2t)^2+w_i^4s^2},
\end{equation}
and in a similar way, we find the expression for the local moments 
\begin{equation}\label{mom-gen-res}
I_q(n) = \frac{1}{(\pi \rho (E)N)^q}\left[\frac{sw_n^2}{(E-d_n-w_n^2t)^2+w_n^4s^2}\right]^q\Gamma (q+1),
\end{equation}
where $\Gamma (z)$ is the gamma function and  $q$ is a positive integer. These two general results allow us to calculate the density of states and the statistics of the eigenvectors for any particular choice of the matrices $W$ and $D$ in Eq.\eqref{model-def}.

\begin{figure}[t]
\centering
\includegraphics[width=\textwidth]{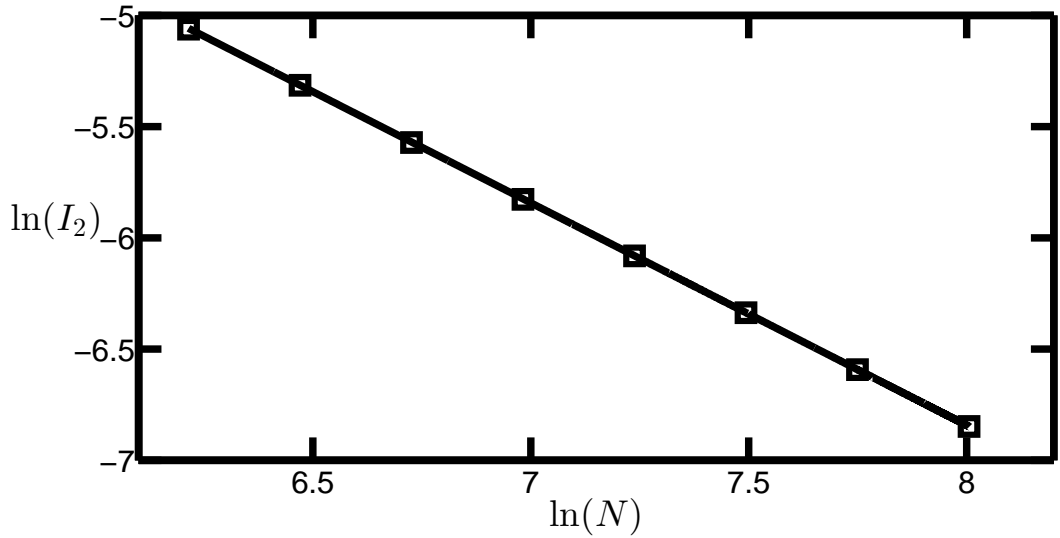}
\caption{The symbols represent the numerical simulation and the solid line is our analytical result. The numerical simulation is over 1000 realizations  for $q=2$ with  $w_i=d_i = N/i$.}\label{fig_lg2}
\end{figure}

Verifying that we recover the GUE case once we set $d_i=0$ and $w_i=1$ is a simple exercise, where we obtain
\begin{equation}
\rho^{GUE}(E)= \frac{1}{\pi} \sqrt{1-(E/2)^2},\ I_q^{GUE}(n) =\frac{ \Gamma (q+1)}{N^q},
\end{equation}
these are the well-known results for the GUE case.

Setting $w_i=1$ we reproduce our previous result derived in Ref. \cite{sumrm}:
\begin{equation}
\begin{aligned}
\rho (E) &= \frac{s}{\pi N}\sum_{i=1}^N \frac{1}{(E-d_i-t)^2+s^2}, \\ I_q (n) &= \frac{1}{(\pi \rho(E)N)^q}\left[\frac{s}{(E-d_n-t)^2+s^2}\right]^q\Gamma (q+1).
\end{aligned}
\end{equation}
At the same time, we can also recover the result  from \cite{R1} by setting $d_i=0$:
\begin{equation}\label{mom_d=0}
\begin{aligned}
\rho (E) &= \frac{s}{\pi N}\sum_{i=1}^N \frac{w_i^2}{(E-w_i^2t)^2+w_i^4s^2}, \\ I_q (n) &= \frac{1}{(\pi \rho(E)N)^q}\left[\frac{sw_n^2}{(E-w_n^2t)^2+w_n^4s^2}\right]^q\Gamma (q+1).
\end{aligned}
\end{equation}

It follows from Eq.\eqref{mom-gen-res} that the scaling of $I_q(n)$ with $N$ remains the same as in the GUE case, provided that $w_i$, $d_i$, $s$ and $t$ are $N$-independent. This implies that the eigenvectors of all such models are extended. Nevertheless their quantitative characteristics, which depend strongly on the ratio $\frac{sw_n^2}{(E-d_n-w_n^2t)^2+w_n^4s^2} $ can change significantly compared to the GUE case. In particular, such eigenvectors can be concentrated on an arbitrarily small fraction of the available space being less ergodic than their GUE counterparts.    

The fact that the local moments $I_q(n)$ depend explicitly only on the corresponding matrix elements $d_n$ and $w_n$ and do not depend on $d_k$ and $w_k$ with $k\neq n$ might be useful for some applications, in which one can control the matrices $D$ and $W$. Indeed, changing the values of $d_n$ and $w_n$ relative to other matrix elements, one can enhance or decrease the corresponding component of the eigenvector in a desirable fashion.  The implicit dependence of  $I_q(k)$ on $d_n$ and $w_n$ with $k\neq n$, which comes from the corresponding dependence of the parameters $s$ and $t$ on all  $d_k$ and $w_k$, can be generally ignored, since the contribution of the term containing  $d_n$ and $w_n$ in Eq.(\ref{sim-eqns}) is by factor $1/N$ smaller than the total contribution of all other terms, unless $E$ is tuned to the resonance value $E_{res}=d_n+w_n^2t$.  

We test our general result by numerical simulations, considering a specific model, in which $w_i = d_i = N/i$. Numerical results for the density of states and the moments of the eigenvectors were produced  by direct matrix diagonalization and they match our analytical expressions with high accuracy.  Fig.~\ref{fig_lg2} shows the results of numerical simulations for $I_2=\sum_n I_2(n)$ with $N$ ranging from $500$ to $3000$ over a total of $1000$ realizations. The eigenvectors that were used in the calculation correspond to the eigenvalues in the vicinity of $E=0$.

%*******************************************************************************************************
%*******************************************************************************************************
%*******************************************************************************************************

\section {Model with random $W$ and $D=0$.} \label{sec_general_formula} 

A particular case of the general model, in which $D=0$ and $W$ is a deterministic matrix was investigated in Ref.\cite{R1}. In this section we study how the results of that work can be generalized to the case of random $W$. Specifically, we focus on the model, in which $w_i$ are independent Gaussian distributed variables with $\avg{w_i}=0$ and $\avg{w_i^2}=\sigma^2$. 

The system of the equations \eqref{mom-gen-res} at $d_i=0$,
\begin{equation}
t = \frac{1}{N} \sum_i^N \frac{w_i^2(E-w_i^2t)}{(E-w_i^2t)^2+s^2w_i^4},\quad \text{and}\quad 1 =\frac{1}{N}\sum_i^N \frac{w_i^4}{(E-w_i^2t)^2+s^2w_i^4},
\end{equation} 
is valid for any particular realization of the random variables $d_i$.  Therefore $s$ and $t$ also become random variables, whose distribution functions can be found by solving  the equations for each realization of $w_i$. As $s$ and $t$ are determined by a large number of independent random variables, they must satisfy some generalization of the law of large numbers and by numerical simulations we infer that the deviation of $s$ and $t$ from their mean values become smaller and smaller as $N \to \infty$. That means that the variables $s$ and $t$ are self-averaging quantities implying that they can be replaced by their mean values $\left<s\right>$ and $\left<t\right>$. Taken this fact into account and averaging  the above equations over $w_i$ we find
\begin{equation}
\left<t\right> = \frac{1}{N}\sum_i^N \left<\frac{w_i^2(E-w_i^2\left<t\right>)}{(E-w_i^2\left<t\right>)^2+\left<s\right>^2w_i^4}\right>,\quad \text{and}\quad 1 =\frac{1}{N}\sum_i^N \left<\frac{w_i^4}{(E-w_i^2\left<t\right>)^2+\left<s\right>^2w_i^4}\right>.
\end{equation}
As $w_i$ are identically distributed, we can simply replace $w_i$ with $x$ and simplify the system to
\begin{equation}\label{system-x}
\left<t\right> = \left<\frac{x^2(E-x^2\left<t\right>)}{(E-x^2\left<t\right>)^2+\left<s\right>^2x^4}\right>_x,\quad \text{and}\quad 1 = \left<\frac{x^4}{(E-x^2\left<t\right>)^2+\left<s\right>^2x^4}\right>_x,
\end{equation}
where $x$ is the Gaussian distributed random variable with $\avg{x}=0$ and $\avg{x^2}=\sigma^2$. 

In order to compute the average of the second equation, we first rearrange its right hand side as follows 
\begin{equation}
%\begin{aligned}
1=\frac{1}{\left<t\right>^2+\left<s\right>^2}\left(1+ \frac{1}{\left<t\right>^2+\left<s\right>^2}\left<\frac{2Ex^2-E^2}{\left(x^2-\frac{E\left<t\right>}{\left<t\right>^2+\left<s\right>^2}\right)^2+\frac{E^2\left<s\right>^2}{\left(\left<t\right>^2+\left<s\right>^2\right)^2}}\right>_x\right),
%\end{aligned}
\end{equation}
The above average over $x$ can be now calculated using the Fourier transform of $P(x) = \frac{1}{\sqrt{2\pi}}\int_{-\infty}^{\infty}d\kappa e^{-\mathrm{i}\kappa x}\hat{P}(\kappa)$, where $\hat{P}(\kappa) = \frac{1}{\sqrt{2\pi}}e^{-\frac{1}{2}\kappa^2 \sigma^2}$:
\begin{equation}\label{avg-x-main}
\begin{aligned}
&\left<\frac{2Ex^2-E^2}{\left(x^2-\frac{E\left<t\right>}{\left<t\right>^2+\left<s\right>^2}\right)^2+\frac{E^2\left<s\right>^2}{\left(\left<t\right>^2+\left<s\right>^2\right)^2}}\right>_x\\ &= \frac{1}{\sqrt{2\pi}} \int_{-\infty}^{\infty}d\kappa \frac{1}{\sqrt{2\pi}}e^{-\frac{1}{2}\kappa^2 \sigma^2}\int_{-\infty}^{\infty}dx \frac{(2Ex^2-E^2)e^{-\mathrm{i}\kappa x}}{\left(x^2-\frac{E\left<t\right>}{\left<t\right>^2+\left<s\right>^2}\right)^2+\frac{E^2\left<s\right>^2}{\left(\left<t\right>^2+\left<s\right>^2\right)^2}}.
\end{aligned}
\end{equation}
Once the integration is completed (see \ref{app_b} for details), we get the expression for the averaged equation
\begin{equation}\label{AvgSys-1}
\begin{aligned}
&1 = \frac{1}{\left<t\right>^2+\left<s\right>^2} \left(1+2\left<t\right>^2+\frac{\mathrm{i}\sqrt{E}}{2}\sqrt{\frac{\pi}{2}}\frac{1}{\sigma}\vast[F_+\left(\left<t\right>,\left<s\right>\right)-F_-\left(\left<t\right>,\left<s\right>\right)\vast]\right),
\end{aligned}
\end{equation}
where we introduced the functions
\begin{equation}
F_{\pm} (x,y) = \frac{e^{-\frac{E(x\pm \mathrm{i}y)}{2(x^2+y^2)\sigma^2}}}{\sqrt{x\mp\mathrm{i}y}}\left(1\pm \mathrm{i}\:\mathrm{erfi}\left[\sqrt{\frac{E(x\pm\mathrm{i}y)}{2(x^2+y^2)\sigma^2}}\right]\right).
\end{equation}
and $\mathrm{erfi}(z)$ stands for the imaginary error function.

\begin{figure}[t]
\centering
\includegraphics[width=\textwidth]{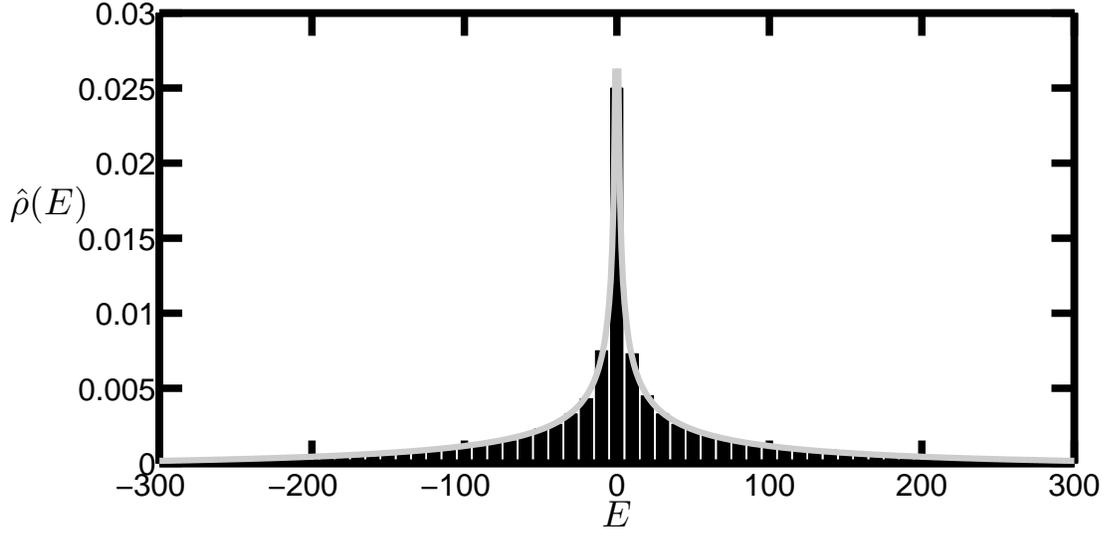}
\caption{The grey line represents the analytical result and the histogram shows the numerical data. The numerical simulations were performed over 1000 realizations of random matrices for $N=1000$ and $\sigma = 10$.}\label{fig_DOS}
\end{figure}

A similar approach is taken to average the first simultaneous equation, which gives
\begin{equation}\label{AvgSys-2}
\begin{aligned}
&\left<t\right> = \frac{\sqrt{E}}{2\left<s\right>}\sqrt{\frac{\pi}{8}}\frac{1}{\sigma} \vast(F_-\left(\left<t\right>,\left<s\right>\right)+F_+\left(\left<t\right>,\left<s\right>\right)\vast).
\end{aligned}
\end{equation}
By solving the system of equations (\ref{AvgSys-1}) and (\ref{AvgSys-2}) numerically, we can find $\left<s\right>$ and $\left<t\right>$ and hence the density of states
\begin{equation}
\hat{\rho} (E) = \frac{2\left<s\right>\left<t\right>}{\pi E}.
\end{equation}

In Fig.~\ref{fig_DOS} we present the results of numerical simulations testing the validity of this expression. One can show that $t\propto \sqrt{E}$ and $s=O(1)$ at $E\to 0$. Therefore the density of states, $ \hat{\rho} (E) \propto  1/\sqrt{E}$, is singular at $E=0$. The origin of this singularity can be understood from the general expression \eqref{dos-main}, according to which the density of states is given by a sum of Lorentzians.  At $d_i=0$ and $E=0$ the contribution of  each of them to $\rho(0)$ has a maximum value proportional to $w_i^{-2}$. Since negative moments are divergent for the Gaussian distribution, the density of states tends to infinity, if $w_i$ are random Gaussian variables.

Employing the same method one can average the expression for the moments of the eigenvectors \eqref{mom_d=0}:
\begin{equation}
\begin{aligned}
\hat{I}_q &\equiv \sum_n^N\left<I_q(n)\right> = \frac{N\left<s\right>^q\Gamma (q+1)}{(\pi \hat{\rho}(E)N)^q}\left<\frac{x^{2q}}{[(E-x^2\left<t\right>)^2+\left<s\right>^2x^4]^q}\right>_x\\&=  \frac{E^q\Gamma (q+1)}{2^q\left<t\right>^qN^{q-1}}\left<\frac{x^{2q}}{[(E-x^2\left<t\right>)^2+\left<s\right>^2x^4]^q}\right>_x.
\end{aligned}
\end{equation}
The calculation of the averaging over $w_i$ can be simplified first by noticing that
\begin{equation}
%\begin{aligned}
\frac{x^{2q}}{[(E-x^2\left<t\right>)^2+\left<s\right>^2x^4]^q} =\frac{1}{(q-1)!}\left[\left(-\frac{1}{2y}\frac{d}{dy}\right)^{q-1}\frac{x^{4-2q}}{(E-x^2\left<t\right>)^2+y^2x^4}\right]_{y=\left<s\right>},
%\end{aligned}
\end{equation}
therefore the averaged  moments of the eigenvectors can be written as 
\begin{equation}
%\begin{aligned}
\hat{I}_q =\frac{qE^q}{2^q\left<t\right>^qN^{q-1}}\left[\left(-\frac{1}{2y}\frac{d}{dy}\right)^{q-1}\left<\frac{x^{4-2q}}{(E-x^2\left<t\right>)^2+y^2x^4}\right>_x\right]_{y=\left<s\right>}.
%\end{aligned}
\end{equation}
The latter average can be evaluated exactly in the same way as one in Eq.\eqref{system-x}. Once the averaging is completed, we arrive at the final result for the moments
\begin{equation}\label{mom-avg-final}
\begin{aligned}
&\hat{I}_q = \frac{q\sqrt{E}}{2^q\left<t\right>^qN^{q-1}}\Vast[\left(-\frac{1}{2y}\frac{d}{dy}\right)^{q-1}\frac{1}{\sigma y}\sqrt{\frac{\pi}{8}}\Vast\{(\left<t\right>+\mathrm{i}y)^{q-1}F_-(\left<t\right>,\left<s\right>)\\&+(\left<t\right>-\mathrm{i}y)^{q-1}F_+(\left<t\right>,\left<s\right>)\Vast\}\Vast]_{y=\left<s\right>}.
\end{aligned}
\end{equation}

\begin{figure}[t]
\centering
\includegraphics[width=\textwidth]{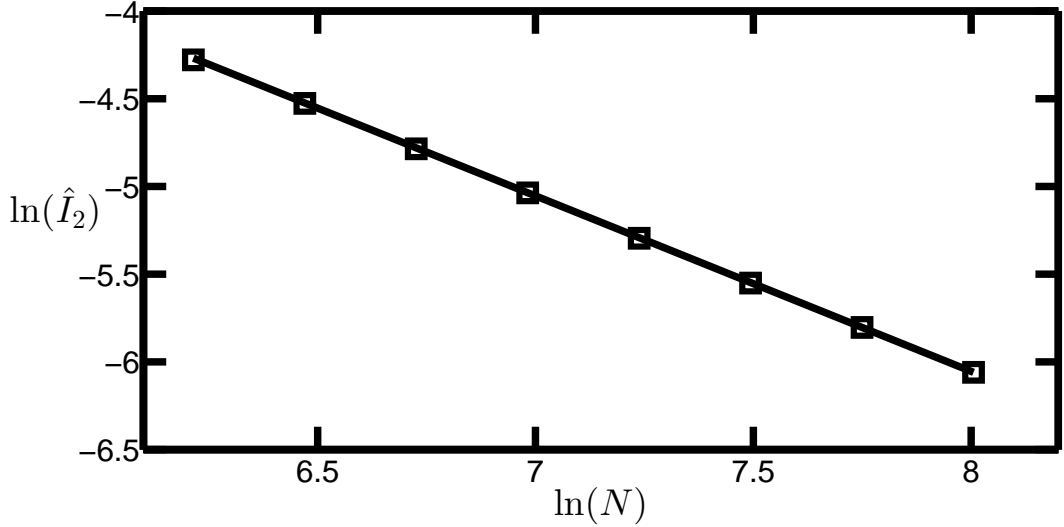}
\caption{The numerical results are given by the symbols and the solid line depicts our analytical result. In this numerical simulation  over 1000 realizations of random matrices, we used $\sigma = 10$.}\label{fig_Iq}
\end{figure}

The derivatives can be calculated explicitly for any integer $q$. Since the final expressions for $\hat{I}_q$ become quite lengthy for higher values of $q$, here we present only an explicit formula for $q=2$:
\begin{equation}
\begin{aligned}
&\hat{I}_2 = \frac{\sqrt{E}}{8N\left<t\right>^2\left<s\right>^2\sigma}\sqrt{\frac{\pi}{2}}\Bigg[\left(\frac{\left<t\right>-\mathrm{i}\left<s\right>}{\left<s\right>}+\frac{\mathrm{i}}{2}\left(1+\frac{E(\left<t\right>+\mathrm{i}\left<s\right>)}{(\left<t\right>^2+\left<s\right>^2)\sigma^2}\right)\right)F_+(\left<t\right>,\left<s\right>)\\& \left(\frac{\left<t\right>+\mathrm{i}\left<s\right>}{\left<s\right>}-\frac{\mathrm{i}}{2}\left(1+\frac{E(\left<t\right>-\mathrm{i}\left<s\right>)}{(\left<t\right>^2+\left<s\right>^2)\sigma^2}\right)\right)F_-(\left<t\right>,\left<s\right>)+\frac{\sqrt{2E}\left<t\right>}{\sqrt{\pi} \sigma (\left<t\right>^2+\left<s\right>^2)}\Bigg].
\end{aligned}
\end{equation}
In order to corroborate the validity of this expression we ran numerical simulations for $\sigma = 10$. The numerical results presented in Fig.~\ref{fig_Iq} along with the analytical solution fully confirm its validity.  The moment with $q=2$ was calculated for  the eigenvectors corresponding the eigenvalues from the vicinity of $E=1$.

According to Eq.\eqref{mom-avg-final} the scaling of $\hat{I}_q \propto N^{1-q}$ is exactly the same as in GUE, indicating that the eigenvectors of this model are qualitatively similar to the GUE eigenvectors. However, if one assumes that $\sigma$ acquires $N$-dependence, then this conclusion cannot be drawn any more. To explore such a possibility, we study the model with $\sigma=N^{\gamma}$, $\gamma>0$. 

\begin{figure}[t]
\centering
\includegraphics[width=\textwidth]{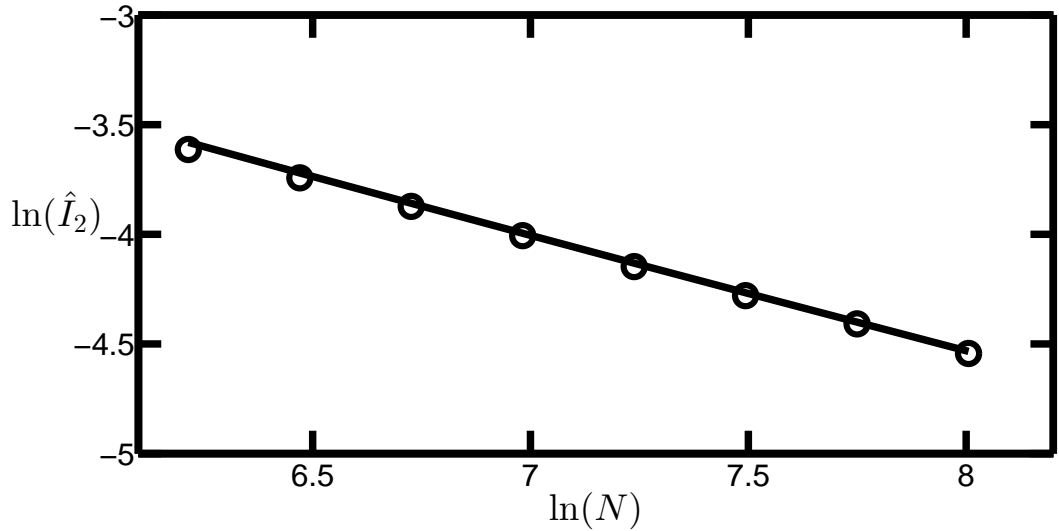}
\caption{This figure shows the results of numerical simulations (symbols) for $\hat{I}_2$ at $\sigma = N^{1/2}$.  The solid line represents our analytical result. The numerical simulations were performed  over 1000 realizations of random matrices. }\label{fig_Iq3}
\end{figure}

Since $\sigma\to \infty$, as $N\to \infty$ we can analyse the asymptotic behaviour of the simultaneous equations when $\sigma \to \infty$, assuming that $E\propto O(1)$, so we set $E=1$ for simplicity. One can show that in this limit  $\left<s\right> \gg \left<t\right>$, therefore we can expand all the expression in  $\left<t\right>/\left<s\right>$ and keep only the leading order terms. Then the asymptotic solution of the simultaneous equations is given by 
\begin{equation}
\left<t\right> \approx \frac{\sqrt{\pi}}{4\sigma}, \quad \left<s\right> \approx 1.
\end{equation}
Substituting this result into the formula for $\hat{I}_q$ we find an asymptotic expression for the moments:
\begin{equation}
\hat{I}_q \approx q\left(\frac{\sigma}{\sqrt{\pi} N}\right)^{q-1}\prod_{k=0}^{q-2} \left|q-\frac{5+4k}{2}\right|
\end{equation}
This result holds for any $\sigma \gg 1$. In particular, for $\sigma=N^{\gamma}$ we have $\hat{I}_q\propto N^{(\gamma-1)(q-1)}$. The scaling of the moments with non-trivial power of $N$ implies that the eigenvectors become fractal in this case with the fractal dimension $D_q=1-\gamma$. There is a clear similarity between this finding and recent results \cite{Kravtsov, sumrm} for non-ergodic states in the Rosenzweig-Porter model \cite{RP60}. Indeed, our results for $\sigma=\mbox{const}$ and  $\sigma=N^{\gamma}$ show that there is a transition at $\gamma=0$ from ergodic to non-ergodic states.  As the exponent $(\gamma-1)(q-1)$ of the scaling law must be negative, we conclude that our result breaks down for $\gamma>1$, where we expect that the eigenvectors become localized. Thus the model we discuss here can be considered as a multiplicative analogue of the Rosenzweig-Porter model. The presence of critical states for the random matrices of the form $W\tilde{H}W$, which do not require any fine-tuning of parameters of a model, might be important for understanding of emergence of such states in various applications such as, for example, critical wave functions of certain biomolecules, reported recently in Ref. \cite{VSCNK15}. 

We  computed $\hat{I}_2$ numerically for $\sigma = N^{1/2}$ for the eigenvectors, whose eigenvalues are sufficiently close to $E=1$, and found the the numerical results are in agreement with our prediction. The corresponding results are given in Fig.~\ref{fig_Iq3}.

%*******************************************************************************************************
%*******************************************************************************************************
%*******************************************************************************************************
\section {Conclusions} \label{sec_conclusions} 

We studied a general class of the structured random matrices given by Eq.\eqref{model-def}. Our main focus was on the statistical properties of the eigenvectors of such random matrices. Using the supersymmetry technique we derived a very general expression for the local moments of the eigenvectors. This result allowed us not only to make predictions about qualitative nature of the eigenvectors, such as a degree of their ergodicity, but also to understand, how particular components of the eigenvectors are affected by the corresponding matrix elements of $W$ and $D$.  

We investigated in detail a special case of the model with $D=0$ and Gaussian distributed $W$. We found that when the variance of $w_i$ scales in a power-law fashion with $N$, the eigenvectors of the model become critical and are characterized by a non-trivial fractal dimension, making such ensemble of random matrices to be similar to the Rosenzweig-Porter model.  

It would be interesting to generalize our results to other random matrix ensembles. Particularly, in many applications instead of the matrix $\tilde{H}$ from the GUE one should deal with matrices from the Gaussian Orthogonal Ensemble or Wishart matrices.  

KT acknowledges support from the Engineering and Physical Sciences Research Council [grant number EP/M5065881/1]. 

\appendix
\section{Pre-exponential factors in Efetov's parametrization}\label{app-parametrization}

The pre-exponential factors calculated by employing Efetov's parametrization are given as follows: 
\begin{eqnarray}
g_{aa}^{BB} &=& \dfrac{E-d_n-w_n^2t+\mathrm{i}sw_n^2\lambda_1 + \mathrm{i}sw_n^2 (\lambda_1 - \lambda_2) \alpha \alpha^*}{(E-d_n-w_n^2t)^2+s^2w_n^4} ,\\
g_{ar}^{BB} &=& -\dfrac{\mu_1sw_n^2\left(1+\dfrac{\alpha \alpha^*}{2}\right)\left(1-\dfrac{\beta \beta^*}{2}\right)+\mu_2^*sw_n^2\alpha^* \beta}{(E-d_n-w_n^2t)^2+s^2w_n^4}, \\
g_{ra}^{BB} &=& -\dfrac{\mu_1^*sw_n^2\left(1-\dfrac{\beta \beta^*}{2}\right)\left(1+\dfrac{\alpha \alpha^*}{2}\right) + \mu_2sw_n^2\beta^*\alpha}{(E-d_n-w_n^2t)^2+s^2w_n^4}, \\
g_{rr}^{BB} &=& \dfrac{E-d_n-w_n^2t - \mathrm{i}sw_n^2 \lambda_1+\mathrm{i}sw_n^2(\lambda_1-\lambda_2)\beta \beta^*}{(E-d_n-w_n^2t)^2+s^2w_n^4}.
\end{eqnarray}

The integration measure reads
\begin{equation}
\mathrm{d}\mu(T) = -\dfrac{\mathrm{d}\lambda_1 \mathrm{d} \lambda_2}{(\lambda_1-\lambda_2)^2} \mathrm{d}\phi_1 \mathrm{d}\phi_2 \mathrm{d} \alpha \mathrm{d} \alpha^*\mathrm{d} \beta \mathrm{d} \beta^*,
\end{equation} 
where $\lambda_1 \in [1,\infty),\ \lambda_2 \in [-1,1],\ \phi_2,\phi_2 \in [0,2\pi]$, and $\alpha, \alpha^*, \beta, \beta^*$ are Grassmann variables, for which the following convention is used  
\begin{equation}
\int \mathrm{d}\alpha\ \alpha =\int \mathrm{d}\alpha^*\ \alpha^* =\int \mathrm{d}\beta\ \beta =\int \mathrm{d}\beta^*\ \beta^* =\frac{1}{\sqrt{2 \pi}}.
\end{equation}
%%%%%%%%%%%%%%%%%%%%%%%%%%%%%%%%%%%%%%%%%%%%%%%%%%%%%%%%

\section{Computing  the average in Eq.\eqref{avg-x-main}}\label{app_b}
The integral over $x$ can be computed by the application of the residue theorem, which gives 
\begin{equation}
\begin{aligned}
&\int_{-\infty}^{\infty}dx \frac{(2Ex^2\left<t\right>-E^2)e^{-\mathrm{i}\kappa x}}{\left(x^2-\frac{E\left<t\right>}{\left<t\right>^2+\left<s\right>^2}\right)^2+\frac{E^2\left<s\right>^2}{\left(\left<t\right>^2+\left<s\right>^2\right)^2}}=\frac{\pi \left(\left<t\right>^2+\left<s\right>^2\right)\sqrt{E}}{2\left<s\right>}\\&\times\left(\frac{\left<t\right>-\mathrm{i}\left<s\right>}{\sqrt{\left<t\right>+\mathrm{i}\left<s\right>}}e^{-\mathrm{i}|\kappa| \sqrt{\frac{E}{\left<t\right>+\mathrm{i}\left<s\right>}}}+\frac{\left<t\right>+\mathrm{i}\left<s\right>}{\sqrt{\left<t\right>-\mathrm{i}\left<s\right>}}e^{\mathrm{i}|\kappa| \sqrt{\frac{E}{\left<t\right>-\left<s\right>}}} \right),
\end{aligned}
\end{equation} 
where we made the following assumptions: $E>0$, $\left<t\right>>0$ and $\left<s\right>>0$.

Therefore the average is equal to
\begin{equation}
\begin{aligned}
&\left<\frac{2Ex^2-E^2}{\left(x^2-\frac{E\left<t\right>}{\left<t\right>^2+\left<s\right>^2}\right)^2+\frac{E^2\left<s\right>^2}{\left(\left<t\right>^2+\left<s\right>^2\right)^2}}\right>_x= \frac{1}{\sqrt{2\pi}} \int_{-\infty}^{\infty}d\kappa \frac{1}{\sqrt{2\pi}}e^{-\frac{1}{2}\kappa^2 \sigma^2}\frac{\pi \left(\left<t\right>^2+\left<s\right>^2\right)\sqrt{E}}{2\left<s\right>}\\&\times\left(\frac{\left<t\right>-\mathrm{i}\left<s\right>}{\sqrt{\left<t\right>+\mathrm{i}\left<s\right>}}e^{-\mathrm{i}|\kappa| \sqrt{\frac{E}{\left<t\right>+\mathrm{i}\left<s\right>}}}+\frac{\left<t\right>+\mathrm{i}\left<s\right>}{\sqrt{\left<t\right>-\mathrm{i}\left<s\right>}}e^{\mathrm{i}|\kappa| \sqrt{\frac{E}{\left<t\right>-\left<s\right>}}} \right),
\end{aligned}
\end{equation}
computing the integral over $\kappa$ we arrive at the result
\begin{equation}
\begin{aligned}
&\left<\frac{2Ex^2-E^2}{\left(x^2-\frac{E\left<t\right>}{\left<t\right>^2+\left<s\right>^2}\right)^2+\frac{E^2\left<s\right>^2}{\left(\left<t\right>^2+\left<s\right>^2\right)^2}}\right>_x = \frac{1}{\left<t\right>^2+\left<s\right>^2}\\&\times \Vvast(1+\frac{\sqrt{E}}{2\left<s\right>}\sqrt{\frac{\pi}{2}}\frac{e^{-\frac{E\left<t\right>}{(\left<t\right>^2+\left<s\right>^2)\sigma^2}}}{\sqrt{\left<t\right>^2+\left<s\right>^2}\sigma}\Vast(e^{\frac{E}{2(\left<t\right>-\mathrm{i}\left<s\right>)\sigma^2}}(\left<t\right>-\mathrm{i}\left<s\right>)^{3/2}\\& \times \left(1-\mathrm{erf}\left[\sqrt{-\frac{E}{2(\left<t\right>+\mathrm{i}\left<s\right>)\sigma^2}}\right]\right)\\&+e^{\frac{E}{2(\left<t\right>+\mathrm{i}\left<s\right>)\sigma^2}}(\left<t\right>+\mathrm{i}\left<s\right>)^{3/2} \left(1-\mathrm{erf}\left[\sqrt{-\frac{E}{2(\left<t\right>-\mathrm{i}\left<s\right>)\sigma^2}}\right]\right)\Vast)\Vvast),
\end{aligned}
\end{equation}
where $\mathrm{erf}$ is the error function.
The expression above can be simplified down to
\begin{equation}
\begin{aligned}
&1 = \frac{1}{\left<t\right>^2+\left<s\right>^2} \Vast(1+\frac{\sqrt{E}}{2\left<s\right>}\sqrt{\frac{\pi}{2}}\frac{e^{-\frac{E\left<t\right>}{(\left<t\right>^2+\left<s\right>^2)\sigma^2}}}{\sqrt{\left<t\right>^2+\left<s\right>^2}\sigma}\Vast(e^{\frac{E}{2(\left<t\right>-\mathrm{i}\left<s\right>)\sigma^2}}(\left<t\right>-\mathrm{i}\left<s\right>)^{3/2}\\& \times \left(1-\mathrm{i}\mathrm{erfi}\left[\sqrt{\frac{E}{2(\left<t\right>+\mathrm{i}\left<s\right>)\sigma^2}}\right]\right)+e^{\frac{E}{2(\left<t\right>+\mathrm{i}\left<s\right>)\sigma^2}}(\left<t\right>+\mathrm{i}\left<s\right>)^{3/2}\\&\times \left(1+\mathrm{i}\mathrm{erfi}\left[\sqrt{\frac{E}{2(\left<t\right>-\mathrm{i}\left<s\right>)\sigma^2}}\right]\right)\Vast)\Vast).
\end{aligned}
\end{equation}
In the integral for the moments we are then able to compute the average by applying the Fourier transform and integrating over the expressions 
\begin{equation}
\begin{aligned}
\frac{1}{\left<t\right>^2+y^2}\left<\frac{(x^2)^{2-q}}{\left(x^2-\frac{E\left<t\right>}{\left<t\right>^2+y^2}\right)^2+\frac{E^2y^2}{(\left<t\right>^2+y^2)^2}}\right>_x  =\frac{1}{\sqrt{2\pi}}\int_{-\infty}^{\infty}d\kappa \frac{1}{\sqrt{2\pi}}e^{-\frac{1}{2}\kappa^2 \sigma^2}\\ \frac{1}{\left<t\right>^2+y^2}\int_{-\infty}^{\infty} dx \frac{(x^2)^{2-q}e^{-\mathrm{i}\kappa x}}{\left(x^2-\frac{E\left<t\right>}{\left<t\right>^2+y^2}\right)^2+\frac{E^2y^2}{(\left<t\right>^2+y^2)^2}}.
\end{aligned}
\end{equation}
Once the integral over $x$ has been completed we arrive at the following
\begin{equation}
\begin{aligned}
&\frac{1}{\left<t\right>^2+y^2}\left<\frac{(x^2)^{2-q}}{\left(x^2-\frac{E\left<t\right>}{\left<t\right>^2+y^2}\right)^2+\frac{E^2y^2}{(\left<t\right>^2+y^2)^2}}\right>_x = \frac{ E^{\frac{1}{2}-q}}{2y}\int_0^{\infty}d\kappa e^{-\frac{1}{2}\kappa^2 \sigma^2}\\& \Bigg((\left<t\right>+\mathrm{i}y)^{q-\frac{3}{2}}e^{-\mathrm{i}\kappa \sqrt{\frac{E}{\left<t\right>+\mathrm{i}y}}}+(\left<t\right>-\mathrm{i}y)^{q-\frac{3}{2}}e^{\mathrm{i}\kappa \sqrt{\frac{E}{\left<t\right>-\mathrm{i}y}}}\Bigg),
\end{aligned}
\end{equation}
the integral over $\kappa$ can also be calculated, which yields the final expression for the moments averaged over $w_i$, this is valid for any integer $q$
\begin{equation}
\begin{aligned}
&\hat{I}_q = \frac{q\sqrt{E}}{2^q\left<t\right>^qN^{q-1}}\Vast[\left(-\frac{1}{2y}\frac{d}{dy}\right)^{q-1}\frac{1}{\sigma y}\sqrt{\frac{\pi}{8}}\Vast\{e^{-\frac{E}{2(\left<t\right>+\mathrm{i}y)\sigma^2}}(\left<t\right>+\mathrm{i}y)^{q-\frac{3}{2}}\\&\Vast(1-\mathrm{i}\mathrm{erfi}\left[\sqrt{\frac{E}{2(\left<t\right>+\mathrm{i}y)\sigma^2}}\right]\Vast)+e^{-\frac{E}{2(\left<t\right>-\mathrm{i}y)\sigma^2}}(\left<t\right>-\mathrm{i}y)^{q-\frac{3}{2}}\\&\Vast(1+\mathrm{i}\mathrm{erfi}\left[\sqrt{\frac{E}{2(\left<t\right>-\mathrm{i}y)\sigma^2}}\right]\Vast)\Vast\}\Vast]_{y=\left<s\right>}.
\end{aligned}
\end{equation}

%*****************************************************************************************************

\end{document}